# Real-time Digital Twins

A TransContinuum Initiative Use Case

By Dirk HARTMANN

We live in a world of exploding complexity driven by technical evolution as well as highly volatile socio-economic environments. Managing complexity is a key issue in everyday decision-making such as providing safe, sustainable, and efficient industrial control solutions as well as solving today's global grand challenges such as the climate change. However, the level of complexity has reached our cognitive capability to take informed decisions. Digital Twins, tightly integrating the real and the digital world, are a key enabler to support decision making for complex systems. They allow informing operational as well as strategic decisions upfront through accepted virtual predictions and optimisations of their real-world counter parts.

**Digital Twins** [6] *are specific virtual representations of physical objects. A Digital Twin integrates all data, models, and other information of a physical asset generated along its life cycle for a dedicated purpose. This is typically reproducing the state and behaviour of the corresponding system as well as predicting and optimising its performance. To this purpose, simulation methods and data-based methods are used.*

Depending on the specific nature, application, and context a wide variety of nomenclature has been introduced, see e.g. [2,4,7,14]. Here, we focus on real-time Digital Twins for online prediction and optimisation of highly dynamic industrial assets and processes. By their nature, Digital Twins integrate and tightly connect several digital key technologies including mathematical modelling, simulation, and optimisation; data analytics, machine learning, and artificial intelligence; data and compute platforms from edge to cloud computing; cybersecurity; human computer interaction; and many more. Only a coordinated research effort as envisaged by the TransContinuum Initiative will allow the realisation of the full potential of Digital Twins - a key tool for decision making addressing today's industrial as well as global challenges.

## Key insights

- Safe, sustainable, and efficient industrial process control and optimisation solutions are (business) critical in today's ever faster, more dynamic, and more volatile industrial environments.
- Digitalisation opportunities in the context of the industrial Internet of Things offer significant potential for novel and more effective control and optimisation concepts.
- The Internet of Things requires novel technologies to overcome today's limitations in terms of data availability in industrial contexts, e.g. due to IP concerns or limited occurrence of failiures, as well as high effort expert-centric serial processes.
- Integrating today's seemingly complementary technologies of model-based and data-based, as well as edge-based and cloud-based approaches has the potential to re-imagine industrial process performance optimisation solutions.

## Key recommendations

- New multi-disciplinary initiatives addressing integrated data- and model-based (first principal physics models-based) approaches beyond the individual silos are required, e.g., investing in the emerging fields of scientific and physics informed machine learning.
- With the increase of pervasive computing power through edge-alising and fast networks, computing becomes more ubiquitous and heterogeneous. An emphasis on corresponding research addressing the efficient exploitation of ubiquitous computing using tailored algorithms as well as workflows is needed.
- The high degree of integration of components and stakeholders in future industrial eco-systems requires research with respect to novel federated integration, workflow, as well as cybersecurity concepts overcoming today's limitations.

## The Industrial Internet of Things and Real-time Digital Twins

Within this contribution, we focus on industrial assets, systems, and processes. In this context, the industrial Internet of Things (IIoT), with its data sensing as well as compute capabilities, offers enormous opportunities. For example, in the context of Industry 4.0 the IIoT is a key driver for novel highly automated manufacturing concepts. Still, many companies are facing challenges to fully exploit opportunities of the IIoT. For example, today only about 30 % of companies capture value from Industry 4.0 solutions at scale [22].

IIoT itself requires vastly different technologies. In the context of social media, online entertainment, online retail, image and video recognition, or natural language procession - domains where data is plentiful and physics-based models do not exist - data analytics and machine learning have been proven extremely powerful technologies [16]. Though the IIoT enables data collection on a scale we could not imagine before, data in industrial contexts is still limited. Data is very specific to the context, e.g., the operational data of an electric drive in a car differs significantly from the data in the context of a compressor, and the different contexts addressed in industrial applications cover an enormous breadth. Therefore, collecting a sufficient amount of data beyond good standard conditions, e.g., for specific failure predictions, is always an enormous challenge. Furthermore, industrial companies often limit data availability outside the company, e.g., due to the risk of reverse engineering of core industrial process know how.

While data might not be available to the extent required for certain machine learning concepts with their big successes in the "Internet of People", fortunately, many industrial systems are well understood in terms of first principles or effective engineering models. Computer simulation is used already since several decades as a decision support tool for research, development, and engineering. These first principle-based physics or engineering models, typically systems of algebraic, ordinary differential or partial differential equations, are often trusted to such an extent that even key validation and verification steps are based on them. Furthermore, the exponential evolutions in terms of computational hardware [15], algorithm capability [13], as well as usability have led today to a point where first principle-based physics or engineering models even allow for real-time prediction capability and thus opening up novel application opportunities beyond R&D. This opportunity will be further accelerated by the concept of executable Digital Twins. These are Digital Twins developed, packaged, and released by simulation experts in R&D departments in combination with the corresponding predictive software functionality such as simulation algorithms. This allows leveraging their prediction capability by anyone at any point of a product's life cycle without the need of additional (simulation) software packages [6] - the ultimate democratisation of Digital Twins beyond their creators and creation tools.

Extending information based on (scarce) data by real-time Digital Twins encapsulating engineering knowledge will be a key success factor for scaling opportunities offered by the IIoT and to tackle new classes of problems in the context of highly optimal industrial complex systems operations.

## Purpose, Business Drivers, and Societal Impact

Potential applications of real-time Digital Twins are plentiful, see Figure 1. Though concepts such as virtual sensors or model predictive control have been around for some time, these technologies have been typically limited by the effort required for manual model building of sufficiently accurate and fast predictive models. Beyond the evolution of hardware and algorithms the relative new research directions integrating physics-based modelling and machine learning [8,16] allows one to overcome these limitations. In combination with other technologies such as edge and cloud computing, AI, 5G and cybersecurity, real-time Digital Twins enable new decision support solutions and paradigms of industrial operation and service with a potential market of around $40 billion by 2026 and compound annual growth rates (CAGR) of 40-60% [10]. This highlights that the IIoT is just taking off with significant socio-economical potential.

# Real-time Digital Twins

| xDT Category | Use this application when you want to… | Selected use cases |
|---|---|---|
| Virtual Testing & Commissioning | … prepare for how your asset or system would interact with other assets, systems, or people. | • Commissioning of automation<br>• Testing of new control strategies<br>• Operator training |
| Virtual Sensing | …measure something in your asset or system, but it isn't feasible to put a sensor there. | • Temperature inside electric rotor<br>• Pressure distribution inside gas turbines<br>• Free-flow inside a sewage network |
| Diagnosis & Identification | …know why your asset or system is behaving the way it is. | • Unbalance detection of large rotors<br>• Leakage detection in a water networks<br>• Predictive maintenance of machines |
| Performance Prediction | …know how your asset or system might behave in future operation. | • Estimating remaining lifetime of motors<br>• Monitoring of coking furnaces<br>• Predicting movement of people |
| Performance Optimization | …inform actions on how to control the asset or system (with or without a Human-in-the-Loop). | • Model predictive control of reactors<br>• Pump schedule optimization<br>• Operating point setting of catalysts |

*Figure 1: Use case categories of real-time Digital Twins. For specific examples see also [6].*

Let us elaborate the potential of addressing safer, more sustainable, energy-aware, and more efficient industrial processes, as summarised in Figure 1, in more detail along two specific use cases:

**Virtual Sensing Increasing Availability of Large Electrical Drives:**

The first use case is real-time prediction of temperatures for rotors of large electrical drives without additional sensor technology[1] (*Figur*e 2). Many electrical drives are temperature limited in terms of their operations. Since effective measurements are often not available, conservative operation schemes are employed limiting availability. Using real-time Digital Twins, virtually sensing temperature fields by combining online data and first-principal models [5], allows for increased availability saving up to 200 000€/h in the oil and gas industry [1]. Pure data-based solutions would not work in this context, since many machines are unique and potential failures during sampling of sufficient data would cost an enormous amount of money.

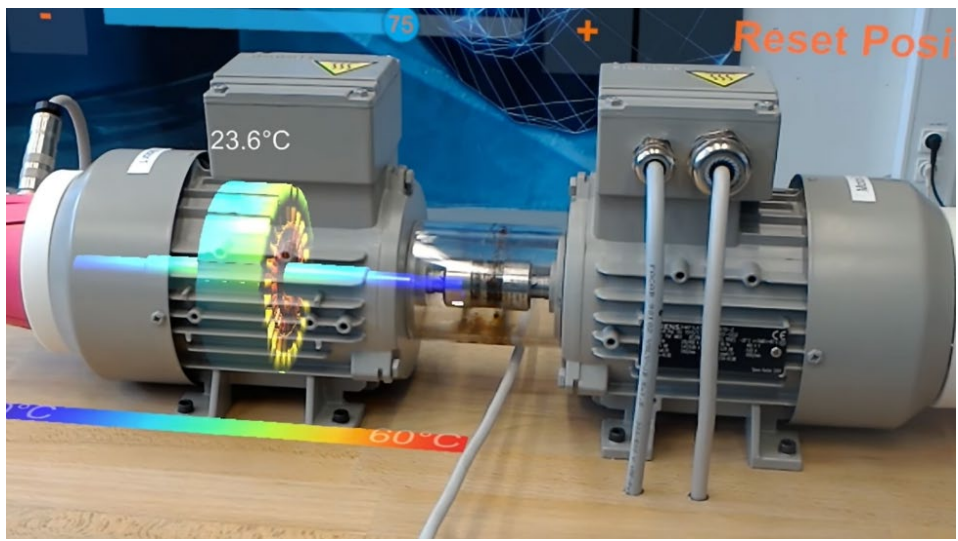

*Figure 2: Virtual sensors based on real-time temperature prediction by Digital Twins in an electrical drive increasing availability.*

---

[1] See also https://youtu.be/86vkjykbHRM
© 2018 Siemens

**Model-Predictive Control Increasing Accuracy in Metal Milling:**

The second use case we would like to highlight are model-based control solutions to increase the accuracy of industrial milling[2] (Figure 3). The accuracy of milling machines is typically limited by the mechanical stiffness of the corresponding machine. Process forces of several hundred Newton lead to deformations of the machine affecting the accuracy of the produced part, e.g., in the context of standard industrial robots, milling process forces could lead to deformations in the range millimetres, which is well above most industrial requirements.

Combining first principal predictions of process forces with a mechanical robot model and online calibration technologies allows the prediction of expected deflections of the robot and to compensate them correspondingly in the control cycle, with update rates on the order of 5 milliseconds [17]. This results in the reduction of machining errors of robots by 90%, sufficient for milling tasks [9]. Classical machine learning approaches will not work since on the one hand metrology in milling machines is very limited (due to dirt, splinter, etc.) and on the other hand combinatorial options in terms of geometries, materials, milling paths, and robot poses would not allow to sample sufficient data.

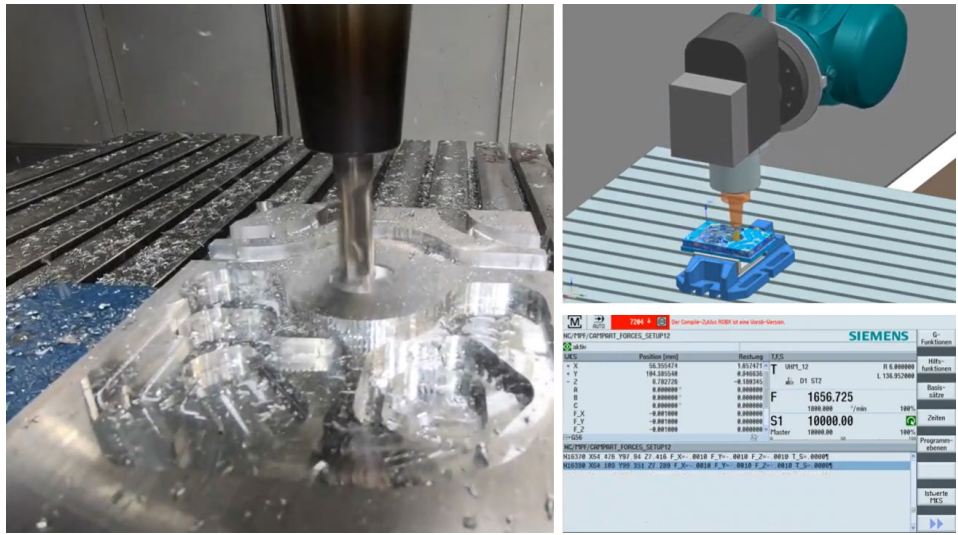

*Figure 3: Digital Twin-based control solution increasing accuracy of milling robots as a key enabler for industrial metal milling.*

## The Workflow and Major Stakeholders

The heart of the Digital Twin concept is its predictive capability based on a seamless integration of various data and model sources. Particularly, this implies first-principle models with different speed and accuracy requirements depending on the specific applications. Today, the generation of corresponding models involves significant manual efforts, which is a major road blocker for many applications. Thus, future highly automated workflows will need to programmatically derive these models from a single source of truth taking the envisaged deployment architectures into account, e.g., start from well verified, validated, and trusted engineering models with the corresponding R&D experts as their champions. This automated transformation of the model will require significant High Throughput Computing (HTC) capability replacing today's manual efforts.

By integrating real-time models with (potentially proprietary) historical or live streaming data, real-time Digital Twins will allow the provision of novel operational solutions. Addressing complex industrial systems will require not only to address the single asset but the overall system or system-of-systems, i.e., addressing the context/system behaviour on different hierarchical levels. This poses enormous challenges to the integration itself. It requires compute power distributed among edge devices, on-premises computing solutions and cloud sources, as well as the secure network communication in between. Furthermore, the operator of the Digital Twin is very likely to be different from the creators of different involved (real-time) Digital Twins as well as the providers of (historical) data. Therefore, the approach requires corresponding intellectual property (IP) protection and trusted data concepts. Additionally, the concept calls for appropriate robustness guarantees and uncertainty quantification of the different components to allow for fail-safe usage by non-experts outside the original authoring tools.

---

[2] See also https://youtu.be/2iIN-9Kno3o (c) 2019 Siemens

Finally, simplified user experience such as low-code approaches will be required to foster industrial adoption.

Today, most solutions are restricted only to well-trained experts limiting scalability.

The corresponding workflows as well as the associated components are depicted in Figure 4.

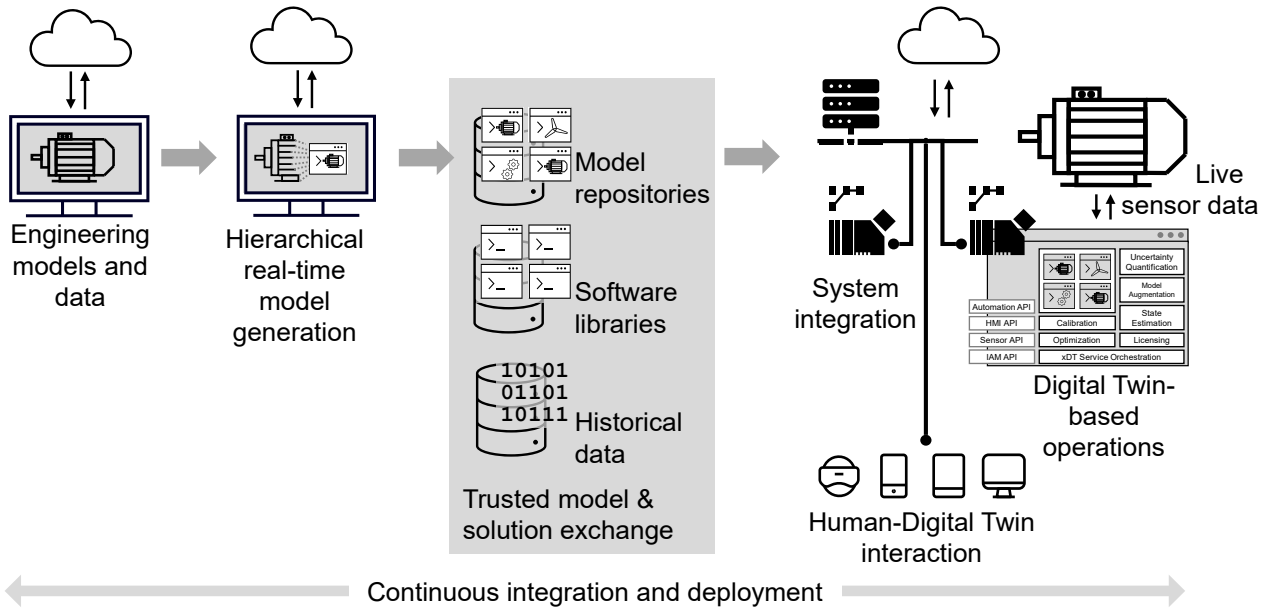

*Figure 4: Selected components of the real-time Digital Twins workflow from the creation to execution in a heterogeneous compute environment combining edge and cloud as well as model- and data-based methods.*

## Challenges along the TransContinuum

First proof of concepts and high manual effort realisations of real-time Digital Twins, such as virtual sensors or model-based controls (see Figure 2 and Figure 3), are available already today. However, scalable approaches are missing. To successfully realise these will require truly multi-disciplinary approaches [19,20] that addresses several challenges across the various technologies in the digital continuum as described below (see also Figure 5)[3]. Beyond these specific technical challenges, we would like to emphasise that adopting Digital Twins will also require rethinking industrial operations and organisations including company processes as well as business models [11]. For example, many of the use cases summarised in Figure 1 clearly favour service-oriented business models [3].

**Mathematical Models, Algorithms and Machine Learning:**

Hierarchical real-time models representing the different components of industrial systems are the main building block of Digital Twins. A key challenge today is to derive corresponding sufficiently accurate real-time models, which is typically a manual task in today's industrial workflows. Linear and non-linear hierarchical model order-reduction technologies offer an opportunity for more automated and thus scalable approaches. Beyond the model creation, the corresponding models must also (self-) calibrate in real-time to the current real-world situation. Since these models will be the basis for decisions, they need to be combined with uncertainty quantification and error estimation technologies to provide accuracy estimates or even guarantees. While the different research fields are established, real-time Digital Twins pose challenges which cannot be solved by today's technology and will require further technology development as well as a tight integration across the different sub-disciplines.

Particularly bringing together the different disciplines of mathematical models, algorithms, and machine learning technologies beyond feeding neural networks with simulation data has the potential to challenge many paradigms in computational science and engineering. First successful attempts can be seen in the newly emerging fields of scientific machine learning, physics-informed neural networks, or neuro differential equations [8,16]. Corresponding approaches will, for example, allow the realisation of very efficient surrogate models or to inform and complement first-principle models with models based on measurements.

---

[3] For a recent technology outlook on Digital Twins, please refer to [12].

**Smart Sensors and Internet of Things:**

The Internet of Things (IoT) provides ubiquitous sensor data availability from the deployed sensors. Still in many industrial applications, sensor and thus data availability is a core issue. Here, real-time Digital Twins can replace missing sensor data by means of physics-based predictions through so called virtual sensors (see Figure 2). These virtual sensors could be used to enrich or extend classical smart sensor applications to make more accurate predictions. Today, the close integration of available sensor technologies and Digital Twins is mostly unexplored. Addressing this will open a fruitful field for exploitation of novel sensing concepts in the context of the IoT.

**Hybrid Compute Infrastructures - From Edge Platforms to (High Throughput) Cloud Computing:**

On the one hand, the preparation of the real-time models requires significant compute capacities. Many methods to reduce model complexity are based on simulations snapshots for different parameter configurations. This exceeds typical workstation capabilities and will foster HTC solutions preferably on cloud infrastructures. On the other hand, advanced networking technologies such as 5G provide enormous ubiquitous and hybrid compute capabilities during industrial operations by combining edge, on premise and cloud computing capabilities including various hardware accelerators. To exploit this opportunity, we require novel concepts for scheduling, resource allocation and novel algorithms cleverly splitting compute tasks in this heterogeneous computing setting. For example, new flexible and adaptable programming and compiler solutions would not require developers to commit upfront to specific architectures and execution models. Thus, the development of high productivity programming environments, performance models and optimisation tools will be crucial.

**Cyber Physical Systems, Smart Networks and Services:**

In terms of developing future physically interactive collaborative systems or more generally Cyber Physical Systems (CPS)[18], Digital Twins will be an imperative technology to connect the real and digital world. With Digital Twins becoming pervasive and particularly in the context of CPS often involving several partners (in an ecosystem), today's very manual model creation and deployment strategies (often tailored for specific monolithic software) will reach their limitations. Novel federated service-oriented architectures will be required to allow for an efficient composition of individual realisations. Furthermore, solutions for seamless data, model, and algorithm transfer across stakeholders are lacking today. Emerging standards, e.g., the Functional Mock Up Interface at the component/interface level, concepts for interdependability modelling and non-functional properties, or concepts like DevOps or MLOps are only the first steps. E.g., the latter require extensions with respect to model-based approaches and product life-cycle management best practices in industrial contexts (Digital Twin Ops).

Beyond the support of cross-ecosystem workflows, novel user interaction technologies will be required to broaden the usage of Digital Twin services and to overcome today's expert-centric workflows. Usability research in this context is in its infancy and increasing usability is seen as a key factor to foster broad industrial adoption. Key technologies which need to be matured are low code approaches in the context of hybrid data- and model-based solutions as well as novel human computer interaction solutions, e.g., leveraging virtual, augmented, or mixed reality solutions.

**Cybersecurity:**

Since Digital Twins will be shared across eco-systems, e.g., suppliers providing models to OEMs, novel cybersecurity concepts are required. Not only do we need mechanisms for appropriate IP and knowledge protection, e.g., prohibiting an infinite execution of models to allow for reverse engineering or for attackers to learn how to interfere with the real system, but there is also the need to ensure that corresponding models are not compromised, e.g., providing traceable and trusted Digital Twins through blockchains. If Digital Twins would provide the wrong information, it could lead to fatal control decisions, e.g., as in the case of the Stuxnet computer worm. At the same time, Digital Twins could be used to check sensor values for plausibility, thus providing novel security solutions to prevent attacks like Stuxnet. Research opportunities at the interface of Digital Twins and Cybersecurity are unexplored today but offer significant potential for safer and more secure industrial operations.

# Real-time Digital Twins

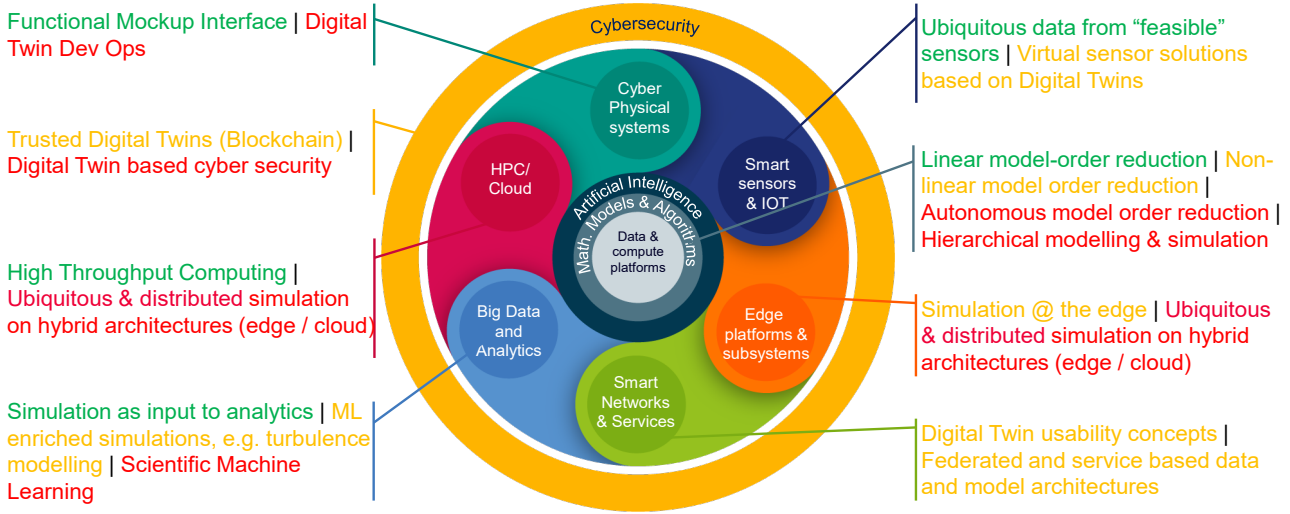

Functional Mockup Interface | Digital Twin Dev Ops

Trusted Digital Twins (Blockchain) | Digital Twin based cyber security

High Throughput Computing | Ubiquitous & distributed simulation on hybrid architectures (edge / cloud)

Simulation as input to analytics | ML enriched simulations, e.g. turbulence modelling | Scientific Machine Learning

Ubiquitous data from "feasible" sensors | Virtual sensor solutions based on Digital Twins

Linear model-order reduction | Non-linear model order reduction | Autonomous model order reduction | Hierarchical modelling & simulation

Simulation @ the edge | Ubiquitous & distributed simulation on hybrid architectures (edge / cloud)

Digital Twin usability concepts | Federated and service based data and model architectures

*Figure 5: Main elements of the digital continuum for the real-time Digital Twins use case including readiness of application and technology (green= established in production, but not optimal in performance; yellow=research, not in production mode; red = novel an*

## Conclusions

Real-time Digital Twins will enable us to rethink paradigms for operation of industrial assets in the context of the IIoT. The concept of real-time Digital Twins integrates the so far opposing algorithmic approaches of first principle and data-based methods as well as the opposed compute paradigms of cloud and edge computing. This allows for overcoming today's data and compute limitations in the context of many industrial systems. Through realising service oriented and federated software frameworks they will enable different parties within eco-systems securely sharing knowledge in the form of data, models and algorithms. Thereby, this will unlock completely new value streams beyond single companies, towards safer, more efficient, energy-aware and scalable industrial operations.

Providing decision support for a safer, accurate, energy-aware, efficient, and sustainable operation of industrial assets and systems, as highlighted by the two use cases presented in *Purpose, Business Drivers and Societal Impact*. Real-time Digital Twins will be a key building block of any industrial operation system of the future. Particularly these concepts build on the historic European strengths in Industry 4.0 as well as computational science and engineering. Therefore, we anticipate that Europe has the potential to play a leading role in the context of real-time Digital Twins.

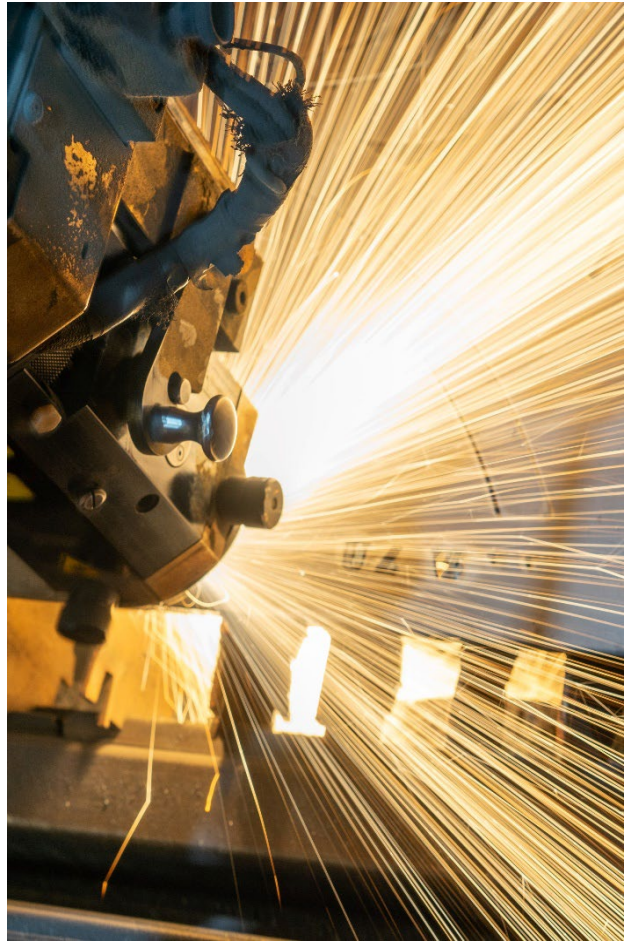

*Dirk Hartmann is Senior principal key research scientist of the technology field "Simulation and Digital Twin" and a Siemens Top Innovatorat Siemens in Munich, Germany.*